\documentclass[11pt,reqno]{amsart}
\usepackage{amsmath, euscript, amssymb, amsfonts}

\theoremstyle{plain}
\newtheorem{theorem}{Theorem}
\newtheorem{lemma}{Lemma}
\newtheorem{corollary}{Corollary}

\theoremstyle{definition}
\newtheorem{definition}{Definition}

\newtheorem*{acknowledgements}{Acknowledgements}

\theoremstyle{remark}

\newcommand{\oR}{{\mathbb R}}

\newcommand{\oC}{{\mathbb C}}

\newcommand{\oK}{{\mathbb K}}

 \newcommand{\supp}{\mathop{\rm supp}\nolimits}

 \newcommand{\R}{\mathop{\rm Re}\nolimits}
 \newcommand{\I}{\mathop{\rm Im}\nolimits}
 \newcommand{\Int}{\mathop{\rm Int}\nolimits}
 \newcommand{\ch}{\mathop{\rm ch}\nolimits}

\begin{document}

 \title{Two classes of generalized functions used in nonlocal
field theory}

\author{M.~A.~Soloviev}
\address{Lebedev
Physical Institute, Russian Academy of Sciences,
 Leninsky
Prospect 53, Moscow 119991, Russia} \email{soloviev@lpi.ru}
\thanks{}

 \begin{abstract}
 We elucidate the relation between the two ways of formulating causality in
 nonlocal quantum field theory:  using analytic test functions
 belonging to the space $S^0$ (which is the Fourier transform of the Schwartz
 space $\mathcal D$) and using test functions in the Gelfand-Shilov spaces
 $S^0_\alpha$. We prove that every functional defined on $S^0$ has the
 same carrier cones as its restrictions to the smaller spaces $S^0_\alpha$.
 As an application of this result, we derive a Paley-Wiener-Schwartz-type
 theorem for arbitrarily singular generalized functions of tempered growth
 and obtain the corresponding extension of Vladimirov's algebra of functions
 holomorphic on a tubular domain.
  \end{abstract}
\maketitle
    \section{Introduction}
   One of the most advanced branches of nonlocal quantum field theory (QFT)
   is based on treating nonlocal fields as highly singular operator-valued
   generalized functions that must be averaged with analytic test functions.
   The Haag-Ruelle scattering theory and the PCT and
   spin-statistics theorems have thus been extended to nonlocal interactions;
      this generalization proved  feasible for an arbitrarily singular
      ultraviolet behavior of the vacuum expectation values (see~\cite{1}).
      Such an approach also turned out to be effective for
       describing  strong interactions phenomenologically~\cite{E}.
       Gelfand and Shilov~\cite{GS} introduced  test
      function spaces of type $S$, which are convenient for these applications.
       The specific choice of test
      functions should be adapted to the model under study, but
      the spaces $S^0_\alpha$ play a dominant role in deriving the general
      theorems  because the Fourier transforms of their elements have compact
      supports and these spaces are hence suitable for fields with
      arbitrarily singular high-energy behavior. The subscript $\alpha$
      controls the decrease of elements of $S^0_\alpha$ at infinity. Namely,
      by definition~\cite{GS},  they decrease no slower than
       $\exp\{-|x/A|^{1/\alpha}\}$ with some $A>0$; this space proves to
      be nontrivial only if  $\alpha>1$. The mathematical foundation for
      extending the results of the axiomatic approach~\cite{SW},~\cite{BLOT} to
      fields defined on $S^0_\alpha$ is their angular localizability~\cite{S0}.
      Specifically, for each  (closed) cone of directions
       $K\subset\oR^d$ we can define the space
      $S^0_\alpha(K)$ in such a way that  $S^0_\alpha(\oR^d)=S^0_\alpha$
      and the correspondence $K\to S^0_\alpha(K)$ is a one-to-one
      mapping satisfying the structural relations
      \begin{equation}
      S^0_\alpha(K_1\cap K_2)= S^0_\alpha(K_1)+
      S^0_\alpha(K_2),\qquad S^0_\alpha(K_1\cup K_2)= S^0_\alpha(K_1)\cap
      S^0_\alpha(K_2).
       \label{0}
      \end{equation}
      Relations~\eqref{0} imply dual relations for the generalized functions
       composing the dual space $S^{\prime\,0}_\alpha$; as a  consequence
       of these dual relations, each element of  $S^{\prime\,0}_\alpha$ has a unique
       minimal closed carrier cone. In the treatment of nonlocal fields of the class
       $S^{\prime\,0}_\alpha$ with a dense invariant domain
        $D$ in the Hilbert space of states, it is natural to replace the
    microcausality axiom with the condition that for any field components
    $\phi_\iota$ and $\phi_{\iota^\prime}$, the matrix elements of
    the    commutator or anticommutator
    \begin{equation}
\langle\Phi,\,
[\phi_{\iota}(x),\phi_{\iota^\prime}(x')]_{\stackrel{-}{(+)}}\Psi\rangle
   \qquad (\Phi, \Psi\in D)
  \label{*}
\end{equation}
are continuous in the topology of $S^0_\alpha(\oK)$, where
$\oK=\{(x,x') \in \oR^8:  (x-x')^2\geq 0\}$. As shown in~\cite{1},
this condition
   ensures the normal   spin-statistics connection and the PCT ivariance of
   nonlocal field theory.

   A number of works (see, e.g., \cite{St},~\cite{L} and the references therein) on
    nonlocal theory use another test function space
    $S^0$, which is just the Fourier transform of the Schwartz space $\mathcal D$
   of smooth functions of compact support.  An argument in favor of this
   choice is that the vacuum expectation values of any  field theory
   (both local and nonlocal) whose state space has a positive metric are
   bounded in any difference variable if the other variables are  fixed
   (this is discussed in great detail in~\cite{W}.)
   For this reason, the space $S^0$, which is a formal limit of
   $S^0_\alpha$ as $\alpha\to\infty$, is apparently suitable for
   realizing the majority of models, although it gives a slightly
   narrower framework.  As shown in~\cite{S2}, here again, there is a natural
   definition of the spaces $S^0(K)$ associated with cones, which leads to an
   alternative formulation of causality as the continuity property of matrix
   elements~\eqref{*} under the topology of $S^0(\oK)$. The connection between
   this formulation and those proposed by other authors is also
   discussed there.

   But a thorough elucidation of the interplay between the two
    generalizations  of local commutativity stated above
   has been  an open issue up to now. The point is
   that the topological structure of the spaces
   $S^0(K)$, which are constructed of Banach spaces using two
   (projective and inductive) limits, is quite complicated. For instance, even
   proving that they are complete is a considerable challenge. In contrast,
   the spaces  $S^0_\alpha(K)$ enter the well-studied class~\cite{K}
   of DFS spaces and have properties that are most convenient for
   applications. In  case of difficulties with functionals and fields
   defined on $S^0$, we may work with their restrictions to $S^0_\alpha$.
   But then the question arises  whether the carrier cones of the
   restrictions defined  using  test functions in
    $S^0_\alpha$ are the same as the carrier cones of the initial functionals.
    As shown in Sec.~3 below,
   the answer is affirmative.  When coupled with formulas~\eqref{0},
   this immediately implies that every functional in $S^{\prime\,0}$ has a
   smallest carrier cone.  Another consequence of this result is established
   in Sec.~5, where the Paley-Wiener-Schwartz theorem, which plays an important
   part in QFT, is extended to the functional class $S^{\prime\,0}$.  It
   should be noted that the direct derivation of this extension by analogy to
   what was done in~\cite{S1} for $S^{\prime\,0}_\alpha$ is vastly more
   sophisticated.  Section~2 is devoted to necessary  preliminaries.
   We do not dwell on the motivation for the  definitions used because it has
   been detailed in~\cite{S0},~\cite{S2}. The main tool used  below to
   derive the theorem on carriers of restrictions is  H\"ormander's
   $L^2$-estimates~\cite{H2}   for solutions of  nonhomogeneous
   Cauchy-Riemann equations. In Sec.~4, this technique is adapted to the
   problem under study, which is required because the weight functions that
   define the norms in H\"ormander's estimates are logarithmically
   plurisubharmonic, whereas
   the indicator functions defining the spaces $S^0(K)$ and $S^0_\alpha(K)$
    are not of such  form.  We refer to Vladimirov's
   treatise~\cite{V} for the facts about plurisubharmonic functions and the
   duality of convex cones.  In Sec.~6, the central theorem of the paper is
   extended to the more general class of spaces $S^0_a$.  Section~7 contains
   concluding remarks.

 \section{Basic definitions and notation}

  \begin{definition} let $U$ be an open cone in $\oR^d$.
 The space  $S^{0,b}(U)$ is the intersection (projective limit)
  of the Hilbert spaces $H^{0,B}_N(U)\quad (B>b,\,\,
N=0,1,2,\dots)$ consisting of entire functions on $\oC^d$ and
endowed with the scalar products
\begin{equation}
 {\langle
f,\,g\rangle}_{U,B,N} = \int \overline{f(z)}
  g(z)\,(1+|x|)^{2N}\,e^{-2B\delta_U(x) - 2B|y|}\,{\rm d}x {\rm d}y \quad
(z=x+iy),
\label{1}
\end{equation}
where $\delta_U(x)$ is the distance of $x$ from
$U$. It can also be defined as the intersection of the Banach spaces
$E^{0,B}_N(U)$  of entire functions with the norms
 \begin{equation}
    \|f\|_{U,B,N}= \sup_{z\in \oC^d}|f(z)| \,(1+|x|)^N\,e^{ - B \delta_U(x) -
    B|y|}.
 \label{2}
\end{equation}
\end{definition}

The equivalence of the system of norms~\eqref{2} to that defined
by  scalar products~\eqref{1} is easily proved from Cauchy's
integral formula in the same way as for the spaces
$S^\beta_\alpha(U)$ in~\cite{S0}. The space $S^0(U)$ is the union
 (inductive limit) of the spaces $S^{0,b}(U)$ ($b\to\infty$).
If $K$ is a closed cone, then $S^0(K)$ is defined as the inductive
limit of the spaces $S^0(U)$, where  $U$ ranges  those open cones
that contain $K$ as a compact subcone, which is denoted by
$K\Subset U$.\footnote{For arbitrary cones $V_1$, $V_2$, the
notation $V_1\Subset V_2$ means that $\Bar V_1\setminus
\{0\}\subset V_2$, where $\Bar V_1$ is the closure of $V_1$.} All
these spaces are continuously embedded in the space $S^0(\{0\})$
of analytic functions of exponential type. It corresponds to the
degenerate closed cone consisting of one point, namely,  the
origin.
 (It is valid and convenient to say that the same space corresponds to the
empty open cone.) The spaces $S^{0,b}(U)$ belong to the class FN
of nuclear Fr\'echet spaces. This fact, which is  essential for applications
to QFT, can be established in the same way as in deriving Theorem 2
 of Ref.~\cite{S1}.  As a consequence, they also belong  to the
Fr\'echet-Schwartz class FS and are Montel spaces. In particular,
they are reflexive. The spaces $S^0(U)$ and $S^0(K)$, being
countable inductive limits of such spaces, inherit nuclearity (see
Sec.~III.7.4 in~\cite{Sch}) and are obviously Hausdorff spaces.
The spaces $S^0(U)$ are complete, as is proved in~\cite{S3}  using
the acyclicity of the injective sequence $S^{0,\nu}(U)$,
$\nu=1,2,\dots$. Together with nuclearity and barrelledness, this
implies that they are  Montel spaces (see Exer.~19 in Chap.~IV
in~\cite{Sch}) and hence are reflexive. Whether $S^0(K)$ has such
properties is still an open question.

 \begin{definition}
 The space  $S^{0}_\alpha(U)$, where $\alpha>1$ and $U$ is an open cone in
 $\oR^d$, is the inductive limit of the Hilbert spaces
  $H^{0,B}_{\alpha, A}(U)$, $A>0$, $B>0$, consisting of entire functions
 on $\oC^d$ and endowed with the scalar products
 \begin{equation}
{\langle f,\,g\rangle}_{U,A,B} = \int\!
\overline{f(z)}\, g(z)\,e^{2(|x/A|^{1/\alpha} - B \delta_U(x) - B|y|)}\,{\rm
d}x {\rm d}y.
   \label{3}
 \end{equation}
\end{definition}

 This inductive limit coincides (see~\cite{S0})  with that of the
Banach spaces  $E^{0,B}_{\alpha,A}(U)$  of entire functions with
the norms
     \begin{equation}
 \|f\|_{U,B,A}= \sup_{z\in \oC^d}|f(z)|
 \,e^{|x/A|^{1/\alpha} - B \delta_U(x) - B|y|} .
    \label{4}
    \end{equation}
The space $S^0_\alpha(K)$, where $K$ is a closed cone,  is defined as the
inductive limit of the spaces $S^0_\alpha(U)$, $U\Supset K$.
All the spaces $S^{0}_\alpha(U)$ and $S^0_\alpha(K)$  are continuously
embedded into $S^0_\alpha(\{0\})$ which is obviously the same as $S^0(\{0\})$.
As shown in~\cite{S1}, they belong to the class  DFS and even to the class
 DFN. (These abbreviations denote the respective strong dual spaces
 of Fr\'echet-Schwartz spaces and of nuclear Fr\'echet spaces.)
 Therefore, they  are  Montel spaces and reflexive.

 \begin{definition}A closed cone $K\subset \oR^d$ is said to be a carrier
 cone of a functional $v\in S^{\prime\, 0}(\oR^d)$ (or $S^{\prime\,
 0}_\alpha(\oR^d)$) if $v$ has a continuous extension to the space
 $S^0(K)$ (or $S^0_\alpha(K)$).
\end{definition}

   Such an extension, if it exists, is unique because
   $S^0_\alpha$ is dense in $S^0_\alpha(K)$, as shown in~\cite{S1}, and this
    also implies  that  $S^0$ is dense in
   $S^0(K)$ (see~\cite{S2} for the details).  Hence, the subspace  of functionals
   carried by $K$ is algebraically identified with $S^{\prime\, 0}(K)$
   or $S^{\prime\, 0}_\alpha(K)$.

  In what follows, we use the following elementary lemma.
   \begin{lemma}
    {\it  Let $E$ be a linear space and let $L_0$, $L_1$, and $L_2$
  be its subspaces endowed with locally convex topologies and such that
   $L_0\subset L_1\cap L_2$. We assume that
  $L_1+L_2$ and $L_1\cap L_2$ are equipped with the respective inductive
   and projective  topologies. If $L_0$ is dense in each of $L_1$, $L_2$,
  and $L_1\cap L_2$  and the injections $L_0\to L_1$, $L_0\to L_2$ are
  continuous, then
  $$
  (L_1+L_2)'=L'_1\cap L'_2,
  $$
  where the dual spaces are  regarded as linear subspaces of
  $L'_0$.}
\end{lemma}

  \begin{proof} We note that $L_0$ is dense in $L_1+L_2$ if it is dense in
   $L_1$ and  $L_2$ and  the natural mapping $(L_1+L_2)'\to L'_0$
   is hence   injective along with $L'_1\to L'_0$ and $L'_2\to L'_0$.
   Clearly, $(L_1+L_2)'\subset L'_1\cap L'_2$ and we need only  show the
   converse inclusion. Let $v\in L'_1\cap L'_2$ and let $v_1$ and $v_2$ be its
   continuous extensions  to $L_1$ and $L_2$.  Because the projective topology on
   $L_1\cap L_2$ is the upper bound of the topologies induced by those of
   $L_1$ and $L_2$, the functionals $v_1$ and $v_2$ are continuous on
   $L_1\cap L_2$ and hence coincide on this subspace by the denseness
   condition. Therefore, the formula $\hat v(f_1+f_2)= v_1(f_1)+ v_2(f_2)$
   defines a linear extension of $v$ to $L_1+L_2$ which is continuous by the
   definition of inductive topology.
\end{proof}

  \section{Theorem on the restriction of functionals}

\medskip
\begin{theorem} Let $v\in S^{\prime\, 0}$ and let
$\alpha>1$.  If the restriction $v|S^0_{\alpha}$ is carried by a
closed cone $K$, then so is $v$.
\end{theorem}
\begin{proof} This statement, combined with the obvious converse
implication, can be expressed by the relation
  \begin{equation}
 S^{\prime\,0}\cap S^{\prime\, 0}_\alpha(K)=S^{\prime\, 0}(K) ,
\label{21}
 \end{equation}
 where all the spaces are regarded as vector subspaces of
 $S^{\prime\, 0}_\alpha$, which is permissible because
$S^0_\alpha$ is obviously dense in $S^0$ and
 $S^0_\alpha$ is dense in $S^0_\alpha(K)$  and in $S^0(K)$ as pointed out
 above.  For $K=\{0\}$, equality~\eqref{21} is valid by definition, and
 the cone  $K$ is henceforth  assumed nontrivial. We  begin by deriving
  the dual formula
  \begin{equation} S^0(K)=S^0+S^0_\alpha(K)
  \label{22}
   \end{equation}
  and then apply Lemma 1. Let $f \in S^{0,b}(U)$, $U\Supset K$.
   Using the dilation invariance of the spaces involved, we
assume that $b<1$  without loss of generality. Here and in the
next two sections, we use the Euclidean norm in
 $\oR^d$, denoted by
$|\cdot|$. We choose a nonnegative function $\chi_0\in C^\infty_0$
with support in the ball $B_\epsilon=\{x\colon|x|<\epsilon\}$ and
such that $\int\chi_0(x)\,{\rm d}x=1$ and set
    $$
   \chi(x)=\int\limits_U\!\chi_0(x-\xi)\,{\rm d}\xi.
   $$
We decompose $f$ as
    $$
   f=f_1+f_2,\quad f_1(z)=f(z)\chi(x), \quad
 f_2(z)=f(z)(1-\chi(x)).
    $$
    The functions $f_1$ and $f_2$  are not analytic  but  respectively
    behave at infinity   as elements of $S^0$ and $S^0_\alpha(K)$.
   Indeed,  we have
 \begin{equation}
 |f_1(z)|\le
 \|f\|_{U,1,N}\,(1+|x|)^{-N}\,e^{\epsilon+ |y|}
  \label{23}
  \end{equation}
  for any  $N$ because $\delta_{U}(x)\le \epsilon$ for $x\in \supp\chi$.
  Further, let  $V$  be an open cone chosen so that
  $K\Subset V\Subset U$.  Then there is a constant
 $\gamma>0$ such that $\delta_{V}(x)\ge \gamma|x|$ for all points of $\supp
 (1-\chi)$ except  a compact subset.  At these points, we have
 $\delta_{U}(x)\le \delta_{V}(x)\le 2  \delta_{V}(x)-\gamma|x|$.
   Therefore,
 \begin{equation}
   |f_2(z)|\le
 C\,\|f\|_{U,1,N}\,e^{-\gamma|x|+2\delta_{V}(x)+|y|}.
\label{24}
\end{equation}
To obtain an analytic decomposition, we write
 \begin{equation}
 f=f'_1+f'_2,\quad  f'_1=f_1-\psi,\quad f'_2=f_2+\psi
 \notag
\end{equation}
and subject $\psi$  to the equations
\begin{equation}
\frac{\partial\psi}{\partial\bar z_j}=\eta_j,
 \label{25}
\end{equation}
where
$$
\eta_j\stackrel{{\rm def}}{=}f\frac{\partial\chi}{\partial\bar
z_j} =\frac{1}{2}f\frac{\partial\chi}{\partial x_j},\quad
j=1,\dots,d.
$$
 The functions $\eta_j(z)$ are nonzero only for $x\in\partial
U+B_\epsilon$, where $\partial U$ is the boundary of $U$, and
satisfy the estimate
\begin{equation}
 |\eta_j(z)|\le C_{j,N}\,(1+|x|)^{-N}\,e^{|y|}
 \label{26}
 \end{equation}
 for each  $N$.
   It remains to verify that there exists a solution of Eqs.~\eqref{25}
  with the required behavior at infinity. We characterize
 this behavior  by a plurisubharmonic function,
 which allows applying an existence theorem due to
 H\"{o}rmander~\cite{H2}.

\begin{lemma}
 Let $U'$ and $U$ be open cones in $\oR^d$ such that $U'\Subset
U$ and let $\alpha>1$. For each system of functions $\eta_j$,
$j=1,\dots,d$, satisfying~\eqref{26} and having support in an
$\epsilon$-neighborhood of the boundary of $U$, there is a
plurisubharmonic function  $\rho(z)$ with values in
$(-\infty,+\infty)$ such that for any $B>\sqrt{d}$ and
$N=0,1,2,\dots$, the following inequalities hold:
  \begin{align}
 \rho(z)&\ge
\max_j\log |\eta_j(z)| &&\text{for}\quad x\in \partial
U+B_\epsilon ,
\label{27}\\
\rho(z)&\le -N\log(1+|x|)  + B|y|+C_{B,N} &&\text{everywhere},
\label{28}\\
\rho(z)&\le-|x|^{1/\alpha} + B|y|+C_B &&\text{for}\quad x\in U',
\label{29}
\end{align}
where $C_{B,N}$ and $C_B$ are constants.
\end{lemma}

 The proof of Lemma 2 is given in the next section, and we now
 derive formula~\eqref{21}.  We choose a cone
  $U'$ such that $V\Subset U'\Subset U$ and set
  $\varrho(z)=2\rho(z)+(d+1)\log(1+|z|^2)$, where $\rho$ is defined
  in  Lemma 2.  In view of~\eqref{27}
   the functions $\eta_j$  belong to $L^2(\oC^d, e^{-\varrho}{\rm
  d}\lambda)$, where ${\rm d}\lambda={\rm d}x{\rm d}y$ is the Lebesgue measure
  on $\oC^d$.  By  the definition of $\eta_j$,   the consistency conditions
  \begin{equation}
  \partial \eta_j/\partial \bar{z}_k =\partial
  \eta_k/\partial \bar{z}_j
  \label{30}
  \end{equation}
are satisfied. Theorem  15.1.2 in \cite{H2} shows that the system
of equations~\eqref{25} has a solution $\psi$ such that
   \begin{equation}
  2\int|\psi|^2
e^{-\varrho}(1+|z|^2)^{-2}{\rm d}\lambda \le \int|\eta|^2 e^{-\varrho}{\rm
d}\lambda.
   \label{32}
  \end{equation}
   It follows that
   $$
   \psi \in
L^2\left(\oC^d, (1+|x|)^{2(N-d-3)} e^{-2B'|y|}{\rm
d}\lambda\right),\qquad \psi \in L^2\left(\oC^d,
e^{2(|x/A|^{1/\alpha}
  -\delta_{V}(x)-B'|y|)}{\rm d}\lambda\right)
   $$
  for all $N$ and for arbitrary $A>1$ and
 $B'>B$. The first membership relation is ensured by
 \eqref{28}, and the second follows from \eqref{29}
 because for $x\not\in U'$, we have $\delta_{V}(x)\ge \gamma'|x|$ with some
   $\gamma'>0$ and obviously $2(|x/A|^{1/\alpha}
   -\delta_{V}(x)-B'|y|)\le -\varrho(z)-2\log(1+|z|^2)$ by
   \eqref{28} again.  Referring to \eqref{23} and \eqref{24} and keeping
 definitions~\eqref{1} and \eqref{3} in mind, we conclude that the
  analytic functions $f'_1$ and $f'_2$ belong to the respective spaces
   $S^{0}$ and $S^{0}_{\alpha}(V)$. Relation~\eqref{22}  is thus proved.

\begin{lemma} The space $S^0_{\alpha}$ is dense in the intersection
  $S^0\cap S^0_{\alpha}(K)$ endowed with the projective topology.
  \end{lemma}

 In fact,  an approximating sequence for
 $f\in S^0\cap S^0_{\alpha}(K)$ is easy to construct by setting
  $f_\nu=\sigma_\nu f$, where $\sigma_\nu(z)$
 is a sequence of Riemann sums for the integral $\int \sigma_0(z-\xi)\,{\rm
 d}\xi$, with $\sigma_0$ being a function in $S^0_{\alpha}$ whose integral is
 unity.   The sequence $f_\nu\in S^0_{\alpha}$ is bounded in both the spaces
    $S^0$ and $S^0_{\alpha}(K)$ and converges to $f$
  uniformly on compact  subsets of $\oC^d$ by the Vitali--Montel theorem.
  Therefore, by the standard argument in Sec.~II.2.1 in~\cite{GS}, $f_\nu\to
  f$ in the topology of either of these spaces.

  We now use Lemma 1, taking into account that the inductive topology
   $\mathcal T$ on $S^0(K)$  defined by the mappings $S^0\to S^0(K)$ and
   $S^0_\alpha(K)\to S^0(K)$ coincides with the original topology $\tau$ of
   this space by  Grothendieck's version~\cite{G} of the open mapping theorem
   (also see  Raikov's Supp.~1 to the Russian edition of~\cite{RR}).
   Indeed, $\tau$ is not stronger than
  $\mathcal T$  by the definition of the latter, and Grothendieck's theorem is
   applicable here because the space $(S^0(K), \tau)$, being an inductive limit
   of  Fr\'echet spaces and a Hausdorff space, belongs to the class $(\beta)$
   (the spaces of this class are also called ultrabornological) and the space
   $(S^0(K), \mathcal T)$ belongs to the class $\mathcal{UF}$, i.e., can be
   covered by a countable family of its Fr\'echet subspaces.\footnote{This
   name is used for a vector subspace of a Hausdorff, locally convex space $E$
   if it can be made into a complete metrizable locally convex space by
   giving a topology that is stronger than the one induced on it by the
   topology of $E$.} This is the case because both the spaces
   $S^0$ and $S^0_\alpha(K)$ are in the class $\mathcal{UF}$  and
    $\mathcal T$ is representable as the quotient topology of the sum
    $S^0\oplus S^0_\alpha(K)$ modulo a closed subspace
   (see  Prop.~28 in Chap.~V in~\cite{RR}  and Lemma 6 in Supp.~1
   to the Russian edition of this book).  Theorem 1 is thus proved.
\end{proof}
\begin{corollary}  For any pair of closed cones $K_1, K_2$ in
$\oR^d$,
 the relation
   \begin{equation}
   S^{\prime\, 0}(K_1\cap K_2)= S^{\prime\, 0}(K_1)\cap
 S^{\prime\, 0}(K_2)
   \label{**}
  \end{equation}
  holds. Each element of $S^{\prime\, 0}$ has a unique minimal carrier cone.
\end{corollary}

   In view of Theorem 1, formula~\eqref{**} immediately follows  from an
   analogous relation obtained in~\cite{S0}  for the spaces
    $S^{\prime\, 0}_\alpha(K)$, which is
   in turn  a consequence of \eqref{0}; we emphasize that its derivation
    essentially uses the fact that these are FS spaces.
   The existence of the smallest carrier cone of  $v\in S^{\prime\, 0}$
    also immediately  follows  from the existence of such
    a carrier for the restriction $v\!\mid\!  S^0_\alpha$ or, alternatively,
    from relation~\eqref{**} by the usual compactness considerations.

 \section{Approximation of the indicator functions \\ by plurisubharmonic functions}

   The possibility of using plurisubharmonic functions to describe
   the topology of spaces occurring in the theory of  Fourier-Laplace
   transformation was discussed in  Sec.~15.2 in \cite{H2}.  In our
   case, where another class of spaces  comes into play, this issue calls
   for further  examination. The method presented below and based
   on  using an analytic function with special properties and on
   systematically constructing the upper envelopes of families of plurisubharmonic
   functions seems quite general and can also be applied to other
   problems.

 \begin{proof}[Proof of Lemma 2]
Let $e$ be a unit vector in $\oR^d$ and $\theta>0$. We let $R_e$
denote the ray $\{\lambda e\colon\, \lambda\ge 0\}$ and
$K_{e,\theta}$ denote the circular cone $\{\lambda x\colon\,
|x-e|\le\theta,\,\,\lambda\ge 0\}$. We assume that  $\theta$ is
less than the angular separation  between the cones $U$ and $U'$.
 It suffices to prove that for every $e$, there exists a plurisubharmonic
 function $\rho_e(z)$  bounded by~\eqref{28} (with a constant
 independent of $e$) and satisfying  estimates of  forms~\eqref{27}
 and \eqref{29} but respectively for $x\in R_e+B_\epsilon$ and $x\not\in
 K_{e,\theta}$.  Then the upper envelope
 \begin{equation}
\rho(z)=\overline{\lim_{z^\prime\rightarrow z}}\,\sup\{\rho_e(z')\colon e\in
\partial U,\, |e|=1\}
  \label{B1}
 \end{equation}
satisfies all the required conditions because
$U'\subset\complement K_{e,\theta}$ for every
 $e\in\partial U$.  We emphasize that according to Sec.~10.3
 in~\cite{V},  function~\eqref{B1} is plurisubharmonic because the family
  $\{\rho_e\}$ is locally uniformly bounded from above.
 The space $S^{\alpha}_0(\oR)$, which is the Fourier transform of
 $S_{\alpha}^0(\oR)$, contains a nonnegative even function  $\omega$
   such that $\supp\omega\subset  [-\delta,\,\delta]$,
   $\int\omega(t)\,{\rm d}t=1$, and $|\omega^{(\nu)}(t)|\le
  C_0 A_0^\nu\nu^{\alpha \nu}$, where $A_0$ and $\delta$ can be taken
arbitrarily small (see  Sec.~IV.8.3 in~\cite{GS}).  Let $\Omega$
be the convolution of $\omega$ by the characteristic function of
the segment   $|t|\le 1+\delta$ and let $1+2\delta<\pi/3$.  Then
$\cos(\xi t)>1/2$ for $|\xi|<1$ and $t\in \supp\Omega$.  We
estimate the  Laplace transform
  $\tilde \Omega(\zeta)$ of $\Omega$ in the strip $|\R \zeta|<1$.
  Taking $\int\Omega(t)\,{\rm d}t=2(1+\delta)$ into account,  we obtain
  \begin{equation}
 |\tilde \Omega(\zeta)|\ge \R \int\!
  e^{it\zeta}\Omega(t)\,{\rm d}t\ge
 \frac{1}{2}\int\limits_{|t|>1}e^{-t\I\zeta}\,\Omega(t)\,{\rm d}t \ge
 \frac{1}{2}e^{|\I \zeta|}\int\limits_{t>1}\Omega(t){\rm d}t=
 \frac{\delta}{2}e^{|\I\zeta|}.
  \label{B2}
 \end{equation}
  Therefore,  the subharmonic function
 $\rho_0(\zeta)=\log(2|\tilde\Omega(\zeta)|/\delta)$ is bounded from below
 by $|\I \zeta|$ in that strip.  According to~\cite{GS} we have $\tilde
  \Omega\in E^{0, 1+2\delta}_{\alpha, A}$, where $A$ is proportional to
  $A_0$,  and hence
  \begin{equation}
   \rho_0(\zeta)\le -|\R \zeta/A|^{1/\alpha}+(1+2\delta)|\I
 \zeta| + \mathrm{Const} .
  \label{B3}
 \end{equation}
  We consider the function
   \begin{equation}
 H(r)=\sup_y\sup_{|x|=r}\max_j\{\log|\eta_j(x,y)|-|y|\}.
  \label{B4}
 \end{equation}
 It follows from \eqref{26} that
  \begin{equation}
   H(r)\le C'_N -N\log(1+r).
  \label{B5}
 \end{equation}

  We can now  construct the functions $\rho_e$.
  We first  assume  that  $e$ is the first basis vector
  and define $\rho_1$ as the upper envelope of the family
 \begin{equation}
  \rho_0(z_1-r)+\sum_{j>1}\rho_0(z_j)+H(r), \quad r\ge 0 .
   \notag
   \end{equation}
   Because $H(r)\not\equiv-\infty$, we have $\rho(z)>-\infty$ everywhere.
   If $x=\R z$ lies in the $\epsilon$-neighborhood of  $\lambda e$,
   then $|x_1-|x||<|x_1-\lambda| +|\lambda-|x||< 2\epsilon$ and
   $|x_j|<\epsilon$ for all $j>1$.  Hereafter, we assume that
   $\epsilon<1/2$.  Then
  $$ \rho_1(z)\ge \sum_j|y_j| +
   H(|x|)\ge |y|+ H(|x|)\ge \max_j\log|\eta_j(x,y)|
 $$
 by our construction, and the required lower bound is  satisfied.
 Further, using~\eqref{B3}, \eqref{B5}, and the elementary inequalities
    $|x_j/A|^{1/\alpha} \ge N\,\log(1+|x_j|)
  -C_{N,A}$, $|x_1|\le |x_1-r| +r $, and $\sum |y_j|\le \sqrt{d}|y|$, we
 conclude that $\rho_1$ satisfies~\eqref{28} if $\delta$ is sufficiently
 small.  Finally, if $x\not\in \pm K_{e,\theta}$, then
 $\sum_{j>1}|x_j|^{1/\alpha}\ge|\theta' x|^{1/\alpha}$ with some
 $\theta'>0$, and if $x\in -K_{e,\theta}$, we have $|x_1-r| \ge |x_1|$.
 Therefore, the last desired bound on $\rho_1$ is also satisfied if $\delta$
 and $A_0$ are sufficiently small. Now let  $e$ be an arbitrary unit vector
 on the boundary of $U$, and let $T_e$ be an orthogonal transformation
 taking it to the first basis vector.  The function $\rho_e(z)= \rho_1(T_ez)$
 is  plurisubharmonic and also satisfies all the required constraints because
 the right-hand sides of \eqref{28} and \eqref{29} are invariant under
 rotations, as is the function $|y|+H(|x|)$ majorizing
 $\max_j\log|\eta_j(x,y)|$.  This completes the proof of Lemma 2.
\end{proof}

 \section{Fourier-Laplace transforms of functionals of the  class
   $S^{\prime\,0}$}

 When coupled with the  Paley-Wiener-Schwartz-type theorem established in
  \cite{S1} for functionals of the class   $S^{\prime\,0}_\alpha$, Theorem 1
  readily implies the following result.

\begin{corollary} Let $V$  be an open connected cone in
  $\oR^d$,
 and let  $\alpha>1$. Suppose a function $\mathbf u(\zeta)$
  is analytic
 on the tubular domain $T^V=\oR^d+iV$ and satisfies the estimate
  \begin{equation}
  |\mathbf u(\zeta)|\,\leq\,C_{\varepsilon,
 R}(W)\, \exp\left\{\varepsilon\, |{\rm Im}\:\zeta |^{-1/(\alpha
 -1)}\right\}\qquad ({\rm Im}\:\zeta\in W,\ |\zeta|\leq R)
  \label{33}
\end{equation}
 for arbitrary $\varepsilon, R >0$  and each cone $W\Subset V$.
 If the boundary value of $\mathbf u$ is a Schwartz distribution
 $($i.e.,  belongs to $\mathcal D')$, then this function satisfies the
 stronger inequality
 \begin{equation}
 |\mathbf u(\zeta)|\leq C_R(W)\, |\I \zeta|^{-N_{R,
 W}}\qquad   (\I \zeta\in W,\, |\zeta|\leq R).
   \label{34}
 \end{equation}
\end{corollary}

In fact, as shown in~\cite{S1}, every function analytic in
 $T^V$ and having  property~\eqref{33}  is the Laplace transform
  of a functional    $v\in S^{\prime\,0}_\alpha(V^*)$, where
    $V^*=\{x\colon  x\eta\ge 0, \forall  \eta\in V\}$ is the dual
    cone\footnote{We note that this cone is closed and convex and even
    properly convex (i.e., contains no entire line) because  $V$ is open.}
    of $V$, and   its boundary value is  the Fourier transform of $v$.
    The Fourier transformation isomorphically maps  $S^0$ onto
    $\mathcal D$ (see Sec.~III.2 in~\cite{GS}), and hence Theorem~1
    immediately gives  $v\in S^{\prime\,0}(V^*)$.
    Now~\eqref{34} results from the elementary estimate
 $$
    |\mathbf u(\zeta)| =|(v,e^{iz\zeta})|\le
    \|v\|_{U,B,N}\|e^{iz\zeta}\|_{U,B,N}
 $$
by  norms~\eqref{2}. Here,  $B$  is arbitrarily large,
     $U$ is any open cone containing
     $V^*$ as a compact subcone, and $N$ depends  on $B$, $U$ in general.
    The cone $U$ and another auxiliary cone $U'$
    should be taken so that $V^*\Subset U\Subset U'\Subset \Int{W}^*$
     (here $\Int{W}^*$ is the interior of ${W}^*$). This
    is possible because $W\Subset V$ implies that $V^*\Subset
    \Int{W}^*$.  Setting $\zeta=\xi+i\eta$, we have
     \begin{equation}
    \|e^{iz\zeta}\|_{U,B,N} = \sup_{x,y}\exp\left\{
   -x\eta-y\xi+N\log\left(1+|x|\right) - B \delta_U(x) - B|y|\right\}.
    \label{35}
     \end{equation}
    Assuming that $|\xi|\le R<B$, we can omit the terms
   dependent on  $y$.  If $x\notin U'$, then $\delta_U(x)>\theta |x|$
   with some $\theta>0$, and for $|\eta|\le R<\theta B$, the exponent
  is dominated by a constant. Finally, let $x\in U'$. The inclusion
   $U'\Subset\Int {W}^*$ implies that there is a $\theta'>0$ such that
$x\eta\ge \theta'|x| |\eta|$ for all $x\in U'$ and $\eta\in W$.
Substituting this inequality in~\eqref{35}, dropping  the term $
\delta_U(x)$, and locating the extremum, we obtain~\eqref{34} with
some $C_R(W)$ proportional to
 $\|v\|_{U,B,N}$.

Another consequence of the same combination of Theorem 1 with Theorem 4
   in~\cite{S1} is an extension of the Paley-Wiener-Schwartz theorem
   to the generalized functions of the class $S^{\prime\,0}$.
We let $\mathcal A_0(V)$ denote the space of functions analytic in
$T^V$  and satisfying \eqref{34} for each $W\Subset V$ and every
$R>0$.  Clearly, it is  an algebra under  pointwise
multiplication.

\begin{theorem}Let $V$ be an open connected cone
  in $\oR^d$ and $V^*$ be its dual cone.
   The Laplace transformation $v\to (v, e^{iz\zeta})$ is an isomorphism of
  the space $S^{\prime\,0}(V^*)$ onto the algebra $\mathcal A_0(V)$.
  Consequently, the elements of $S^{\prime\,0}$ with a given closed
  carrier cone compose a convolution algebra.
   The function $\mathbf u(\zeta)$ in
  $\mathcal A_0(V)$ that is the Laplace transform  of
  a functional $v\in S^{\prime\,0}(K)$ tends to its Fourier transform
  $\tilde v$ in the topology of $\mathcal D'$ as $\I \zeta\to 0$ inside a
  fixed cone $W\Subset V$.
\end{theorem}

\begin{proof} We have just seen that every functional belonging to
  $S^{\prime\, 0}(V^*)$ has a Laplace transform defined on
   $T^V$ and satisfying~\eqref{34}. Applied to its restriction
    to $S^0_\alpha$, Theorem~4 in~\cite{S1} shows that the
   Laplace transform is analytic in this tube.
  As noted above, by the same theorem, every function analytic in
   $T^V$ and having  property~\eqref{33}  (and  particularly
   ${\bf u}\in \mathcal A_0(V)$) is the Laplace transform of a
   certain $v\in S^{\prime\, 0}_\alpha(V^*)$, and
   $\int{\bf u}(\xi+i\eta)f(\xi){\rm d}\xi\to (\tilde v,f)$ as $W\ni\eta\to
   0$ for each $f\in S^\alpha_0$. On the other hand, it is well known that
   every function analytic on $T^V$ and satisfying~\eqref{34}
  has a boundary value belonging to
   $\mathcal D'$, which is zero only if the function vanishes
    (see Theorem 3.1.15 in~\cite{H1}).  Therefore, $v$ belongs to
   $S^{\prime\,0}$ and, by Theorem 1, to the space $S^{\prime\,
   0}(V^*)$ as well.  Hence, the Laplace transformation is a one-to-one mapping
  of $S^{\prime\,0}(V^*)$ onto $\mathcal A_0(V)$.
  The weak convergence  ${\bf u}(\xi+i\eta)\to \tilde v$ on elements of
  $\mathcal D$ implies the convergence in the strong topology of $\mathcal
  D'$ because it is a Montel space.
  Furthermore, Theorem 4 of~\cite{S1}  shows that
  $S^{\prime\,0}_\alpha(V^*)$ is a convolution algebra and the convolution
   $v_1*v_2$ of its two elements corresponds to the
  product ${\bf u_1}\cdot{\bf u_2}$ of their Laplace transforms.
  If $v_1, v_2\in S^{\prime\,0}(V^*)$, then ${\bf u_1}\cdot{\bf u_2}\in
  \mathcal A_0(V)$ and hence $v_1*v_2\in S^{\prime\,0}(V^*)$. Therefore,
  $S^{\prime\,0}(V^*)$ is a convolution algebra. It is worth noting that
  an arbitrary closed properly convex cone
  $K\subset\oR^d$ is the dual cone of the interior of $K^*$.
  Theorem 2 is thus proved.  We also note that  Theorem~2 and the
  relation $V^*=(\ch V)^*$, where $\ch$ signifies the convex
  hull, imply that $\mathcal   A_0(V)=\mathcal A_0(\ch V)$.
\end{proof}

  The algebra $\mathcal A_0(V)$  is a generalization of Vladimirov's
  algebra~\cite{V},~\cite{V2}, formed by the Laplace transforms of  tempered
  distributions that have support in a closed properly convex cone, to the
  case of analytic functionals with an arbitrary singularity, which means an
  arbitrary fast increase of the Laplace transforms at infinity.
  The algebra $\mathcal A_0(V)$ can be made into a topological algebra
  by regarding it as the projective limit of the family of spaces
   $A_{0,r}(W)$ ($r>0$, $W\Subset V$) defined in turn as the inductive limits
  of the Banach spaces $A_{0,R, N}(U)$
  ($R>r$, $N=0,1,2,\dots$, $U\Supset W$) whose elements are analytic on
   $T^U_R=\{\zeta\in \oC^d\colon |\zeta|<R,\,\, \I\zeta\in U\}$, with
  $U$ an open cone, and have the finite norm
   \begin{equation}
  \|u\|_{U,R,N} = \sup_{\zeta\in T^U_R}|\I\zeta|^N |u(\zeta)|.  \label{36}
  \end{equation}
   The spaces $A_{0,r}(W)$ belong to the class DFS because the natural
  injections  $A_{0,R', N'}(U') \to A_{0,R, N}(U)$
  ($R'>R, N'<N, U'\Supset U$) are compact mappings, i.e., the image  of the
  unit ball of  $A_{0,R', N'}(U')$ is relatively compact in  $A_{0,R,
  N}(U)$.  In fact, by the Montel theorem, we can choose a sequence  $u_\nu$
  from any infinite  subset of this image such that it converges to an
  analytic function $u$ uniformly on each compact set in $T^{U'}_{R'}$.
  In particular, for every $\epsilon>0$, we have
  $$
   \sup_{\zeta\in \overline{T^U_R},\,
  |\I\zeta|\geq\epsilon}|\I\zeta|^N |u-u_\nu|\leq\epsilon
   $$
  if $\nu$ is sufficiently large. On the other hand, $\|u\|_{U',R',N'}\leq 1$
  and hence
   $$
   \sup_{\zeta\in
    T^U_R,\,|\I\zeta|\leq\epsilon}|\I\zeta|^N |u-u_\nu|\leq\epsilon\,
    \|u-u_\nu\|_{U',R',N'}\leq 2\epsilon.
   $$
  Therefore, $u_\nu\to u$ in the norm $\|\cdot\|_{U,R,N}$.
  The Laplace transformation $S^{\prime\,0}(V^*)
    \to \mathcal A_0(V)$ is continuous in the strong topology of
    $S^{\prime\,0}(V^*)$.  Moreover, it is continuous even if this space is
    given the projective limit topology by the natural embeddings in the
     DFS spaces $S^{\prime\, 0,b}(U)$, $b>0$, $U\Supset
    V^*$.  This is the case because the constant $C_R(W)$ in~\eqref{34}
    can be chosen proportional to $\|v\|_{U,B,N}$, as shown above.
    We note that according to Sec.~IV.4.5 in~\cite{Sch},
    the projective limit topology is consistent  with the duality of
     $S^0(V^*)$ and $S^{\prime\,0}(V^*)$.

     In the simplest, but important, case where  $V$ is the semi-axis
     $\oR_+$ and $V^*=\Bar{\oR}_+$, the space
    $S^{\prime\,0}(\Bar{\oR}_+)$ coincides with $S^{\prime\,0}(\oR_+)$, and we
    can be assured that its strong topology is identical to the projective
    limit topology because of the regularity of the injective sequence of
    spaces $S^{0,\nu}(\oR_+)$.  This property, which means that every bounded
    set in $S^0(\oR_+)$ is contained and bounded in some
     $S^{0,\nu}(\oR_+)$,  is in turn ensured by the acyclicity of the
    sequence (see~\cite{P}). We can also assert that $S^{\prime\,0}(\oR_+)$
     endowed with the strong topology is an ultrabornological space and
    belongs to the class $\mathcal{PUF}$.  The last conclusion
     can be deduced in the same manner as an analogous statement
    for $\mathcal D'$ in Supp.~2 to the Russian edition
    of~\cite{RR} because $S^0(\oR_+)$ is complete and belongs to Grothendieck's
    class $(S)$. Applying  Raikov's generalization of the open mapping
    theorem, which is proved in the same place, we then infer that the strong
    topology of $S^{\prime\,0}(\oR_+)$ coincides with the bornological
    topology associated with the inverse image of the topology of
      $\mathcal A_0(\oR_+)$.  Hence, the inverse image of every bounded set of
      $\mathcal A_0(\oR_+)$ is strongly  bounded in
      $S^{\prime\,0}(\oR_+)$.  In other words, in this case the Laplace
      transformation is not only an algebraic but also a bornological
      isomorphism. This consideration demonstrates the subtleties arising
      in dealing with  analytic functionals of the class
      $S^{\prime\,0}$.  In contrast, in the case of DFN
      spaces $S^{\prime\,0}_\alpha$, the ordinary open mapping theorem
    immediately shows that the Laplace transformation
     $S^{\prime\,0}_\alpha(V^*)\to \mathcal A_0^\alpha(V)$ is a topological
    isomorphism (see~\cite{S1}).

  \section{Generalization to the spaces $S^0_a$}

     In some instances, there is a need  to use more general
    spaces $S^0_a$ instead of $S^0_\alpha$.  We recall~\cite{GS} that $S^0_a$ is
    specified by a sequence of positive numbers $\{a_\nu\}$ and consists of
     the smooth functions  on $\oR^d$ that satisfy the inequalities
     \begin{equation}
     |x^k\partial^{\,q}f(x)|\leq C A^{|k|}
    B^{|q|} a_{|k|},
     \label{37}
      \end{equation}
     where  $C$, $A$, and $B$ are constants dependent on $f$
     and the usual multi-index notation is used.
   The behavior of $f\in S^0_a$ as $|x|\to\infty$ is described by the
    function $1/\mathbf{a}(|x/A|)$, where $|x|=\max_j|x_j|$ and
      $$
     \mathbf{a}(r)\stackrel{{\rm def}}{=}\sup_\nu\frac{r^\nu}{a_\nu}.
      $$
        The function $\mathbf{a}(r)$ is convex and monotone increasing faster
        than any power of its argument. We can assume, without loss of
     generality, that the sequence $\{a_\nu\}$ also satisfies some regularity
     conditions, namely,
       \begin{equation}
      a_0=1,\qquad a_{\nu+1}\geq a_\nu,\qquad
     a_\nu^2\leq a_{\nu-1} a_{\nu+1}.
      \label{38}
     \end{equation}
     The first  is a normalization and the others are features of the
     regularized sequence
     $a_\nu^*=\sup_{r\geq 1}r^\nu/\mathbf{a}(r)$ which determines
      the same indicator function (for $r\geq 1$) because $a_\nu^* \leq
     a_\nu$ and $a_\nu^* \geq r^\nu/\mathbf{a}(r)$. The definition of $S^0_a$
     also includes the requirement
     \begin{equation}
     a_{\nu+1}\leq Ch^\nu
    a_\nu, \label{39}
    \end{equation}
     where $C$ and $h$ are constants, which ensures that the multiplication by
    $x_j$ does not take the functions out of the space.
     Under the same condition,\footnote{Some additional
    restrictions were also imposed in~\cite{GS}, but they are unnecessary. A simple
    construction of the theory of Fourier transformation for  spaces
    of the  type $S$ using only~\eqref{38}  is given in~\cite{S5}.} the
    Fourier operator maps $S^0_a$ isomorphically onto the space  $S_0^a$. That
    space coincides with the space $\mathcal D^{\{a_\nu\}}$ used by
    Roumieu~\cite{R} as the basic space for the theory of ultradistributions.
    (these are just the elements of the dual space $\mathcal
    D^{\prime\,\{a_\nu\}}$). The space $S_0^a$ is nontrivial if and only if
    it satisfies the nonquasianalyticity criterion (see Sec.~1.3 in
    \cite{H1}),
    which is
     \begin{equation}
     \sum_{\nu=1}^\infty     a_\nu^{-1/\nu}<\infty
     \label{40}
     \end{equation}
     in Mandelbrojt's   version.
    From~\eqref{40} coupled with~\eqref{38}, it follows that
    \begin{equation}
     \mathbf{a}(r)\leq C_\epsilon    e^{\epsilon r},
     \label{41}
      \end{equation}
     where $\epsilon$ can be taken arbitrarily small.
  In fact, \eqref{38} means that $\log a_\nu$ is a convex function of the
  index and increases monotonically beginning with the value zero.
  Therefore, the terms of series~\eqref{40} is monotone decreasing.
  In particular,
  $$
  \sum\limits_{\mu>\nu/2}^\nu a_\mu^{-1/\mu}\geq
  (\nu/2)a_\nu^{-1/\nu},
  $$
  and the convergence of the series implies that
   $\nu^\nu\leq\epsilon^\nu a_\nu$ for sufficiently large $\nu$.
   Clearly, each element of $S^0_a$  has an analytic continuation to $\oC^d$
   satisfying
      \begin{equation}
      |f(x+iy)|\leq C\,e^{-\log \mathbf{a}(|x/A|)+
  B\sum |y_j|}.
      \label{42}
      \end{equation}
      Applying Cauchy's formula and
  using~\eqref{41} and the inequality \begin{equation}
     \mathbf{a}((1-\lambda)r_1+\lambda r_2)\leq \mathbf{a}(r_1)
 \mathbf{a}(r_1)
     \label{43}
     \end{equation}
     which follows from the convexity and monotonicity of
  $\mathbf{a}(r)$ (with the normalization $\mathbf{a}(0)=1$) and holds for
     each $\lambda\in [0,1]$, we readily see that, conversely,
      \eqref{42} implies  estimate~\eqref{37} for the derivatives of the
     restriction of the entire function $f$ to $\oR^d$ with constants $A$ and $B$
     which are  generally different from those in~\eqref{42} but can be taken
     arbitrarily close to them.  Therefore, we can use inequality~\eqref{42}
     as the basis for the definition of $S^0_a$ and then define
      $H^{0,B}_{a,A}(U)$, $E^{0,B}_{a,A}(U)$ and also
     $S^0_a(U)$, $S^0_a(K)$ using the replacement
      \begin{equation}
     |x/A|^{1/\alpha}\,\, \longmapsto \,\,\log \mathbf{a}(|x/A|)
     \label{44}
     \end{equation}
     in formulas~\eqref{3} and \eqref{4}.  The notion of a carrier
      cone thus naturally extends to these spaces.
    It is significant that $S^0_a(U)$ and $S^0_a(K)$ are  DFN spaces
     if~\eqref{39} is fulfilled.  This can be shown in a way analogous to
     that used in~\cite{S1} for $S^0_\alpha(U)$, the only difference being that the
     canonical injection  $H^{0,B}_{\alpha, A}(U)\to H^{0,B'}_{\alpha,
     A'}(U)$ is a nuclear mapping for arbitrary $A'>A,\,B'>B$, whereas an
     analogous statement for $H^{0,B}_{a, A}(U)$ is true if $A'$ and
     $B'$ are sufficiently large  compared with $A$ and  $B$, which has
  no effect on the ultimate conclusion.  Thereafter, the proof of
     Theorem~1 extends to the spaces $S^0_a$ almost literally,
     with replacement~\eqref{44} in~\eqref{29} and further. By Theorem 1.3.5
     in~\cite{H1}, condition~\eqref{40} ensures that
      $S^a_0(\oR)$ contains a nonnegative even function $\omega$ supported
      in an arbitrarily small segment $[-\delta,\,\delta]$ and having the
      properties
      \begin{equation}
  \int\omega(t)\,{\rm d}t=1, \qquad |\omega^{(\nu)}(t)|\le C_0 A_0^\nu a_\nu,
 \label{45}
      \end{equation}
  where $A_0$  is sufficiently large compared with $1/\delta$.
   But according to~\cite{R} (Lemma~1 in Chap.~II), for each sequence
  $a_\nu$ satisfying~\eqref{38} and~\eqref{40},
    there is a sequence $a'_\nu$ satisfying the same conditions and such that
     $a'_\nu\leq C_\varepsilon \varepsilon^\nu a_\nu$ for any
    $\varepsilon>0$. Therefore, $A_0$ in~\eqref{45}  can be taken arbitrarily
    small (with a proper increase of $C_0$), which allows
    constructing
    an analogue of the function $\rho_0$ used to derive Lemma~2.
    To ensure that the corresponding analogue of $\rho_1$ has the desired
    properties, we can use the inequality
     $$\sum_{j=1}^d \log
    \mathbf{a}(|x_j|) \geq\log \mathbf{a}(|x/d|),
    $$
     which, as well
     as~\eqref{43},  follows from the regularity properties of $\mathbf{a}$.
     Ultimately, we obtain  the following generalization of Theorem~1.

\begin{theorem}  The restriction of $v\in S^{\prime\, 0}$
to each nontrivial space $S^0_a$ specified by a sequence
  $a_\nu$ satisfying \eqref{38} and \eqref{39} has the same carrier cones as
 the functional $v$ itself.
 \end{theorem}

 \section{Conclusion}
     In~\cite{St}, which was devoted to constructing the scattering theory for
     nonlocal fields defined on the space $S^0$ (denoted by $Z$ there) and,
   in particular, to deriving an analogue of the
   Lehmann-Symanzik-Zimmermann reduction formulas, the locality axiom was
   replaced by some regularity properties of Green's functions in momentum
   space.  The results obtained here can be applied to clarify the
   relation between Steinmann's conditions and the above-stated
   generalization of local commutativity which is closer to the intuitive
   idea of causality and, as already noted, ensures the normal
   spin-statistics relation and the existence of the PCT symmetry.
   A tentative conclusion is that they are actually equivalent.
   Because  $S^0$ is a basic space of functional analysis, a deeper insight
   into the topological properties of the spaces
    $S^0(K)$ and $S^{\prime\,0}(K)$, including a closer examination of
   possibilities for  extending the results concerning $S^{\prime\,0}(\oR_+)$
   that are stated at the end of Sec.~5, is also  mathematically interesting.
   But a major aim in this paper is to show how we can obviate the
   topological problems arising in concrete applications by using the
   restrictions of functionals to appropriate test function spaces
   with more convenient properties. We also  note that as shown
    in~\cite{S4},  an analogue of Theorem 1 holds for the restrictions of
   elements of  $S^{\prime\,0}_\alpha$ to smaller spaces of the same class.

   \begin{acknowledgements} The author thanks A.~Smirnov for useful discussions.
   This work was supported
 in part by the Russian Foundation for Basic Research (Grant No.~05-01-01049)
  and the Program for Supporting Leading Scientific Schools (Grant
 No.~LSS-1578.2003.2).
\end{acknowledgements}



\begin{thebibliography}{99}


 \bibitem{1}  M.~A.~Soloviev, {\it Theor. Math. Phys.,} \textbf{121}
(1999) 1377.

 \bibitem{E}  G.~V.~Efimov, {\it Problems in  Quantum Theory of Nonlocal
Interactions} [in Russian],  Nauka, Moscow, 1985.

 \bibitem{GS}   I.~M.~Gelfand and G.~E.~Shilov,
 {\it Generalized Functions}, Vol. 2,  Acad. Press, New York, 1968.

  \bibitem{SW}  R.~F.~Streater and A.~S.~Wightman,
             {\it PCT, Spin and Statistics and All That},
             Redwood City, Addison-Wesley, 1989.

 \bibitem{BLOT} N.~N.~Bogoliubov, A.~A.~Logunov, A.~I.~Oksak, and I.~T.~Todorov,
{\it General Principles of Quantum Field Theory}, Kluwer,
Dordrecht, 1990.

\bibitem{S0}   M.~A.~Soloviev, {\it Lett. Math. Phys.,} \textbf{33} (1995)
49.

 \bibitem{St}  O.~Steinmann, {\it Commun. Math. Phys.,} \textbf{18} (1970)
 179.

 \bibitem{L}   W.~L\"ucke, {\it J. Math. Phys.},  \textbf{27} (1986)
 1901.

 \bibitem{W}   A.~S.~Wightman, {\it Introduction to some aspects of the
 relativistic dynamics of quantized fields}, in: ``High Energy Electromagnetic
 Interactions and Field Theory,'' Carg\'ese Lectures in Physics, 1964
 (M.~Levy, ed.) pp.~171-291, Gordon \& Breach, New York, 1967.

 \bibitem{S2} M.~A.~Soloviev, {\it J. Math. Phys.}, \textbf{39} (1998)
 2635.

 \bibitem{K}  H.~Komatsu, {\it J. Math. Soc. Japan.},  \textbf{19} (1967)
 366.

 \bibitem{S1} M.~A.~Soloviev, {\it Commun. Math. Phys.}, \textbf{184}
(1997) 579 [arXiv:hep-th/9601005].

 \bibitem{H2}   L.~H\"ormander, {\it The Analysis of Linear Partial
 Differential Operators}, Vol.~2,  Springer,  Berlin, 1983.

 \bibitem{V}  V.~S.~Vladimirov, {\it Methods of the Theory of Functions of Many
Complex Variables}, MIT, Cambridge, Mass., 1966.

 \bibitem{Sch} H.~H.~Schaefer, {\it Topological Vector Spaces},
 MacMillan, New York, 1966.

 \bibitem{S3}  M.~A.~Soloviev, {\it Theor. Math. Phys.}, \textbf{128}
(2001) 1252 [arXiv:math-ph/0112052].

 \bibitem{G}  A.~Grothendieck,   {\it Produits tensoriels topologiques et
espaces nucl\'earies} (Mem. Amer. Math. Soc., Vol.~16),
Providence, R.I., 1955.

\bibitem{RR}  A.~P.~Robertson and W.~Robertson, {\it Topological Vector
Spaces},   Cambridge Univ. Press,  Cambridge, 1964.

   \bibitem{H1}     L.~H\"ormander, {\it The Analysis of Linear Partial
 Differential Operators}, Vol.~1,
 Springer,  Berlin, 1983.

   \bibitem{V2}  V.~S.~Vladimirov, {\it  Generalized Functions in
Mathematical Physics},  Mir,  Moscow, 1979.

  \bibitem{P} V.~P.~Palamodov, {\it Russ. Math. Surv.}, \textbf{26} (1971)
  1.

 \bibitem{S5} M.~A.~Soloviev, {\it Theor. Math. Phys.}, \textbf{52}
(1982) 854.
 \bibitem{R}  C.~Roumieu, {\it  Ann. Sci. \'Ecole Norm. Sup.},  \textbf{77}
  (1960) 41.

 \bibitem{S4} M.~A.~Soloviev, {\it J. Math. Phys.},  \textbf{45} (2004)
 1944.


\end{thebibliography}
\end{document}